\documentstyle[preprint,aps,epsfig]{revtex} % for PREPRINT

\tighten       %Gives single-space (RevTex 3.0)

\begin{document}
\draft

\title{Axisymmetric pulse recycling and motion in bulk semiconductors}
\author{ L. L. Bonilla, R. Escobedo}
\address{Universidad Carlos III de Madrid, Escuela Polit\'{e}cnica
Superior, 28911 Legan\'{e}s, Spain  }

\author{F. J. Higuera}
\address{E.T.S.\ Ingenieros Aeron\'auticos,
       Pza. Cardenal Cisneros 3, 28040 Madrid, Spain}

\date{\today}

\maketitle

\begin{abstract}
The Kroemer model for the Gunn effect in a circular geometry (Corbino disks)
has been numerically solved. The results have been interpreted by means of
asymptotic calculations. Above a certain onset dc voltage bias,
axisymmetric pulses of the electric field are periodically shed by an
inner circular cathode. These pulses decay as they move towards the outer
anode, which they may not reach. As a pulse advances, the external current
increases continuously until a new pulse is generated. Then the current
abruptly decreases, in agreement with existing experimental results.
Depending on the bias, more complex patterns with multiple pulse shedding
are possible.
\end{abstract}

%%%%%%%%%%%%%%%%%%%%%
%\begin{multicols}{2}
%%%%%%%%%%%%%%%%%%%%%
\narrowtext

\section{Introduction}
\label{sec:introduction}
Propagation of pulses naturally occur in excitable media, that exhibit a
large response when a sufficiently strong perturbation disturbs the only
stable stationary homogeneous state \cite{mur93}. Examples are the
propagation of an action potential along the axon of a nerve \cite{kee98},
the propagation of a grass fire on a prairie, pulse propagation through
cardiac cells \cite{kee98}, reaction-diffusion \cite{kur84} or ecological
systems \cite{mur93}. While a vast literature is devoted to the
mathematical description of pulses propagating in unbounded media, less
is known about pulse generation from boundaries and propagation in finite
domains, particularly in multidimensional domains.

That boundaries and boundary condition play an important role in pulse
creation and annihilation is well understood in the subject of
semiconductor instabilities \cite{nie95}. Device geometry and bias
conditions are crucial for instabilities and related nonlinear dynamics to
appear. Nice examples can be found in the experiments by Willing and Maan
on repeated pulse propagation in semi-insulating GaAs \cite{wil94}. They
considered rectangular samples with two attached parallel planar contacts
or with point contacts at different dc voltage. In the first situation,
planar pulses were periodically generated at the cathode and moved towards
the anode where they disappeared. This phenomenon is analogous to the
well-known Gunn effect in bulk n-GaAs \cite{gun63}. In the case of point
contacts, circular waves were repeatedly generated at the cathode and
vanished before arriving at the anode \cite{wil94}. Theoretical studies of
self-sustained oscillations in semi-insulating GaAs are scarce even in
one-dimensional geometries (cf.\  Ref.\ \onlinecite{bon99} and references
cited therein). However, the observed phenomena can be qualitatively
understood within the simpler Kroemer model for the Gunn effect in bulk
n-GaAs \cite{kro66}. An asymptotic study of this model on a one-dimensional
spatial support can be used to understand pulse propagation in samples with
planar contacts \cite{bon97}. A simple study of pulse propagation in
samples with point contacts could consist of analyzing the Kroemer model
in a axisymmetric sample: a circular sample of bulk n-GaAs with a point
contact (cathode) at its center and an attached concentric circular outer
contact (anode). This configuration is known as {\em Corbino geometry}
\cite{nie95}. In this paper, we carry out a numerical study of pulse
dynamics in the Kroemer model with Corbino geometry. Depending on the dc
voltage bias and contact resistivity, we observe stationary field and
current and self-sustained oscillations (periodic or not) due to pulse
propagation and recycling. Pulses may or may not arrive at the anode
before a new pulse is generated at the cathode. These results are
presented in Section \ref{sec:3} after a short description of the Kroemer
model in Corbino geometry is given in Section \ref{sec:2}. The numerical
results are interpreted by means of an asymptotic analysis in Section
\ref{sec:4}. Section \ref{sec:5} contains our conclusions and the
numerical method we use is described in Appendix \ref{appendix}.

\section{Equations and boundary conditions}
\label{sec:2}

\begin{center}
\begin{figure}[ht]
\epsfxsize=70mm
\epsfysize=45mm
\epsfbox{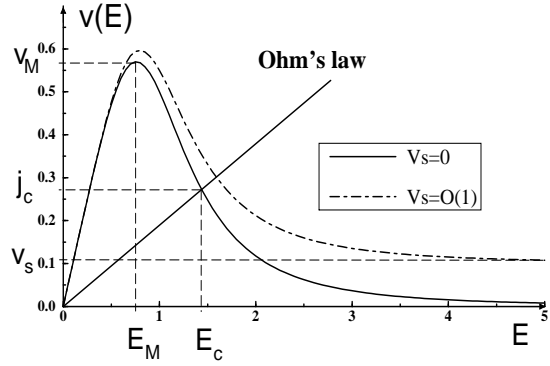}
\vspace{0.2cm}
\caption{Drift velocities and Ohm's law.
$v(E) = |\vec{v}(\vec{E})|$ has a maximum $v_{M} = 3^{3/4}/4$ at
$E = E_{M} = 1/3^{1/4}$ (for $v_s=0$), followed by a region of negative
differential mobility for $E>E_{M}$. At large fields $E\gg 1$, the
electron velocity monotonically decreases to a value $v_s$, which may
be zero.}
\label{velocities}
\end{figure}
\end{center}

The Kroemer model consists of the following equations and boundary
conditions (in dimensionless units) for the concentration of free
carriers (electrons), $n$, and the electric potential, $\varphi$:
\begin{eqnarray}
\frac{\partial n}{\partial t} + \nabla \cdot (n \vec{v}
- \delta \nabla n) = 0, \label{1}\\
\nabla ^2 \varphi = n - 1, \label{2}\\
\vec{v}(\vec{E}) = \vec{E}\, \frac{1+ v_{s} E^{3}}{1+E^{4}}\,,
\label{3}\\
\vec{x} \in \Sigma_{c} : \enspace
\vec{E} \cdot \vec{N} = \rho\, (n \vec{v} - \delta \nabla n) \cdot
\vec{N} \quad \mbox{ and } \quad \varphi = 0,
\label{4}\\
\vec{x} \in \Sigma_{a} : \enspace \vec{E} \cdot \vec{N} = \rho\, (n
\vec{v} - \delta \nabla n) \cdot \vec{N} \quad \mbox{ and } \quad
\varphi = \Phi. \label{5}
\end{eqnarray}
Here (\ref{1}) and (\ref{2}) are the charge continuity and Poisson
equations, respectively. The dimensionless electric field is $\vec{E} =
\nabla \varphi$ and $E = |\vec{E}|$. In these equations, the electron
density has been scaled with the uniform concentration of donor impurities
in the semiconductor, $N_D= 10^{15}$ cm$^{-3}$, and the electric field with
the field characterizing the intervalley transfer responsible for the
negative differential mobility involved in the Gunn oscillation, $E_R=3.1$
kV/cm. Distances and times have been measured with the dielectric length
and the dielectric relaxation time, $l_1 =\epsilon E_{R}/(e N_{D})
\approx 0.276\mu$m, $l_1/(\mu_0 E_R)\approx 1.02$ ps, respectively
($\mu_0$ is the zero-field electron mobility; see, {\it e.g.},
\cite{hig92} for details). The unit of electric potential is $E_R
l_1\approx 0.011$ V. The carrier drift velocity of Eq.\ (\ref{3}),
$\vec{v}(\vec{E})$, is already written in dimensionless units, and it has
been depicted in Fig.\ \ref{velocities}. We assume that the diffusion
coefficient is constant, $\delta\approx 0.013$ (at 20K).

%\vspace{-.2cm}
%\begin{center}
%\begin{figure}[ht]
%\epsfxsize=70mm
%\epsfysize=45mm
%\epsfbox{figure1.eps}
%\vspace{0.2cm}
%\caption{Drift velocities and Ohm's law.
%$v(E) = |\vec{v}(\vec{E})|$ has a maximum $v_{M} = 3^{3/4}/4$ at
%$E = E_{M} = 1/3^{1/4}$ (for $v_s=0$), followed by a region of negative
%differential mobility for $E>E_{M}$. At large fields $E\gg 1$, the
%electron velocity monotonically decreases to a value $v_s$, which may
%be zero.}
%\label{velocities}
%\end{figure}
%\end{center}

Boundary and bias conditions need to be imposed at the interfaces between
semiconductor and contacts, $\Sigma_{c,a}$. Our boundary conditions
(\ref{4}) and (\ref{5}) assume that the normal components of electron
current density and electric field are proportional at the
semiconductor--contact boundary (Ohm's law) \cite{hig92}, (in these
equations, $\vec{N}$ is the unit normal to $\Sigma_{c,a}$, directed
towards the semiconductor). For simplicity, we choose all contact
resistivities $\rho$ to be equal. Bias conditions are chosen to be
$\varphi=0$ at the cathode $\Sigma_{c}$ (injecting contact) and
$\varphi=\Phi$ (the applied voltage) at the anode $\Sigma_{a}$ (receiving
contact). Typically $\delta >0$ is very small, so that diffusion matters
only inside boundary layers near the contacts or inside thin shock
waves~\cite{hig92,bon97}. The latter are charge accumulations that will be
treated simply as discontinuities of the electric field~\cite{bon97,hig92}.
Thus diffusion effects may be left out of the conservation equation
(\ref{1}) when interpreting the results. If we set $\delta = 0$, the first
boundary condition in (\ref{5}) should be omitted.

We can write an Amp\`ere's equation for the total current density
(electronic plus displacement), $\vec{j}$, by eliminating $n$ from (\ref{1})
using (\ref{2}):
\begin{eqnarray}
\nabla \cdot \vec{j} = 0, \quad\quad\quad \mbox{ with } \nonumber\\
\vec{j} = (1+\nabla ^2 \varphi ) \vec{v} - \delta\, \nabla
(\nabla^2 \varphi) + \frac{\partial \vec{E}}{\partial t} \,.
\label{6}
\end{eqnarray}
In the Corbino geometry considered in this paper, this equation can be
simplified further. Let $r_{c}$ and $r_a>r_c$ be the radii of cathode and
anode, respectively. Electric field and current density are now directed
along the radial direction, $\vec{E}= E(r,t) \vec{r}/r$, $E(r,t) = \partial
\varphi(r,t)/\partial r$ and $\vec{j} = J(t)\vec{r}/r^2$, so that (\ref{6})
becomes
\begin{equation}
\frac{\partial E}{\partial t} + v(E) \left[ {1 + \frac{1}{r}
\frac{\partial (r E)}{\partial r} } \right]
- \delta\, \frac{\partial }{\partial r} \left[ \frac{1}{r}
\frac{\partial (r E)}{\partial r} \right] = \frac{J}{r}
\, , \label{7}
\end{equation}
where $2 \pi J(t)$ is the current through the
external circuit, $i(t)= \int_{\Sigma_{c}} \vec{j}\cdot\vec{N}\, dA= 2 \pi
J(t)$. Equation~(\ref{7}) for $E(r,t)$ and $J(t)$ should be solved with the
following bias and boundary conditions
\begin{eqnarray}
{1\over L}\,\int_{r_{c}}^{r_{a}} E \, dr = \phi, \hspace{3.1cm}\label{8}\\
E = \rho \left( { \frac{J}{r} - \frac{\partial E}{\partial t} } \right)
\quad \mbox{ at } \quad r = r_{c},\, r_{a}, \label{9}
\end{eqnarray}
where $L\equiv r_a - r_c$ and $\phi = \Phi/L$.

It is known (see, {\it e.g.}, \cite{hig92}) that planar dipole waves may
appear in long samples when $\rho > 4/3$, for which the straight line $j=
E/\rho$ representing Ohm's law intersects the drift velocity curve $j=v(E)$
at a point $(E_{c},j_{c})$ on the second branch of this curve, as
sketched in Fig.~\ref{velocities}. (In particular $E_{c}\approx \rho^{1/4}$
and $j_{c} \approx \rho^{-3/4}$ for large $\rho$). This is the case
we will consider in the present paper.

\section{Numerical results}
\label{sec:3}
\begin{figure}[ht]
\begin{center}
\epsfxsize=75mm
\epsfysize=60mm
\epsfbox{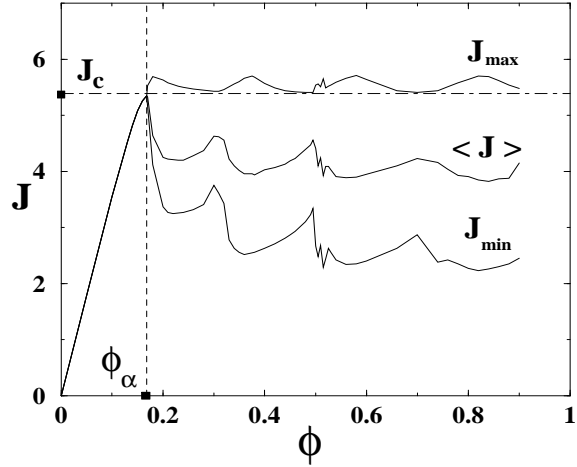}
\vspace{0.4cm}
\caption{Current-voltage characteristic curve $J(\phi)$ for $v_s=0$. If $0<
\phi< \phi_{\alpha}$, the stable electric field profile is stationary. For
$\phi >\phi_{\alpha}$, we have depicted the maxima, minima and
time-averaged value of the current self-oscillations. There are
small-amplitude current self-oscillations if $\phi_{\alpha}<\phi <
2\phi_{\alpha}/3$, and large-amplitude self-oscillations due to pulse
recycling and motion if $2\phi_{\alpha}/3<\phi<\phi_{\omega})$.
$\phi_{\omega}$ is finite for $v_s>0$ and infinite for $v_s=0$.}
\label{J-FI_0}
\end{center}
\end{figure}

We have solved numerically Equations (\ref{7}) and (\ref{8}) together
with the boundary conditions (\ref{9}) at $r=r_c$ and $r=r_a$. Appropriate
initial conditions were given for $E(r,0)$. Parameter values were $\rho =
2$, $\delta = 0.013$, $r_c =10$ and $r_a = 50$ and $90$ (i.e. $L=40$ and
$80$ resp.). The bias $\phi$ was used as a control parameter for two
different electron velocity curves $v_s=0$ and $v_s = 0.1$, representing
zero and nonzero saturation velocities at high electric fields. These
parameter values are appropriate for n-GaAs and consistent with previous
studies~\cite{hig92}. Consider firstly the characteristic current--voltage
curve $J(\phi)$ of Fig.\ \ref{J-FI_0}. There we can mark three different
regimes, already present in experiments \cite{wil94}.

%
%\begin{figure}[ht]
%\begin{center}
%\epsfxsize=75mm
%\epsfysize=60mm
%\epsfbox{figure2.eps}
%\vspace{0.4cm}
%\caption{Current-voltage characteristic curve $J(\phi)$ for $v_s=0$. If $0<
%\phi< \phi_{\alpha}$, the stable electric field profile is stationary. For
%$\phi >\phi_{\alpha}$, we have depicted the maxima, minima and
%time-averaged value of the current self-oscillations. There are
%small-amplitude current self-oscillations if $\phi_{\alpha}<\phi <
%2\phi_{\alpha}/3$, and large-amplitude self-oscillations due to pulse
%recycling and motion if $2\phi_{\alpha}/3<\phi<\phi_{\omega})$.
%$\phi_{\omega}$ is finite for $v_s>0$ and infinite for $v_s=0$.}
%\label{J-FI_0}
%\end{center}
%\end{figure}
%\vspace{-0.4cm}

\textbf{Regime I: } $0<\phi<\phi_{\alpha}\approx 0.168$. Stable solutions
are stationary and $J(\phi)$ is well approximated by a straight line with
slope $\approx 31.5$.

\begin{figure}[ht]
\begin{center}
\epsfxsize=80mm
\epsfbox{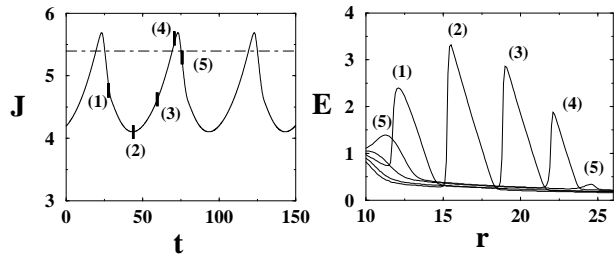}
\vspace{0.2cm}
\caption{Total current density (left) and electric field profiles (right)
for $\phi = 0.18$. The electric field profiles are depicted at the times
marked on the graph of $J(t)$. The horizontal line in the latter corresponds
to the value $J_c$. }
\label{smallosc}
\end{center}
\end{figure}

\textbf{Regime II: } $\phi_{\alpha}<\phi<3\phi_{\alpha}/2 \approx 0.25$.
Above the onset bias for current oscillations, there are small-amplitude
(10-20\% of the overall current signal) sinusoidal current
self-oscillations. The oscillation maxima and minima are about $J_c
\approx 5.4$ and $J_{min}\approx 4$, respectively. The electric field
profile is a triangular pulse which is recycled at the cathode, advances
and it soon disappears at $r\approx 25$ (quenched-mode oscillation). See
Fig.\ \ref{smallosc}.

%\vspace{.2cm}
%\begin{figure}[ht]
%\begin{center}
%\epsfxsize=80mm
%\epsfbox{figure3.eps}
%\vspace{0.2cm}
%\caption{Total current density (left) and electric field profiles (right)
%for $\phi = 0.18$. The electric field profiles are depicted at the times
%marked on the graph of $J(t)$. }
%\label{smallosc}
%\end{center}
%\end{figure}

\begin{center}
\begin{figure}[ht]
\epsfxsize=85mm
\epsfbox{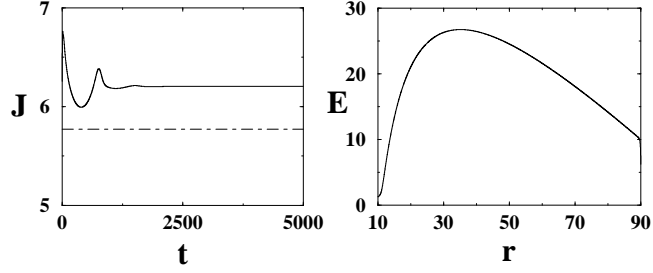}
\vspace{0.2cm}
\caption{Stationary solution for $\phi =20$, $L=80$ and $v_s=0.1$. $J$
evolves towards $J\approx 6.2$ greater than $J_c=5.77$. The maximum
electric field, $E_+ \approx 26.75$, is reached at $R_b \approx 35.12$.
Field values at the boundaries are $E(r_c)=1.25$ and $E(r_a)=0.14$.}
\label{fi20}
\end{figure}
\end{center}

\textbf{Regime III: } $3\phi_{\alpha}/2<\phi<\phi_{\omega}$. The upper
critical bias $\phi_{\omega}$ is finite for $v_s>0$ (e.g.\ $\phi_{\omega}
\approx 15$ if $v_s=0.1$) and infinite for $v_s=0$. There are
large-amplitude (60\% of the overall current signal) current
self-oscillations. These oscillations are mostly time-periodic, although
there are narrow bias intervals of aperiodic oscillations. Their maxima
$J_{max}(\phi)$ are always close to $J_c$, while their minima
$J_{min}(\phi)$ take values on a wider range of currents. Electric field
profiles consist of moving triangular pulses. For $\phi>\phi_{\omega}$, the
stable field profile is again stationary as shown in Fig.\ \ref{fi20}.
These results qualitatively agree with the experimental observations of
self-sustained oscillations in semi-insulating GaAs reported in Ref.\
\cite{wil94}. Experimental samples were rectangular and contained two
well-separated point contacts. It was observed that self-oscillations of
the current were due to circular dipole waves that were recycled at the
cathode, expanded towards the anode and vanished without ever reaching
it. The current signal was similar to that in Fig.\ \ref{valleys} below.
Semi-insulating GaAs is described by model equations different from the
Kroemer model. However, qualitative agreement of experimental data with
our results for the axisymmetric Kroemer model suggest that a theoretical
interpretation of self-oscillations similar to that in Section
\ref{sec:4} could also be appropriate for semi-insulating GaAs. 
%
%\vspace{-.5cm}
%\begin{center}
%\begin{figure}[ht]
%\epsfxsize=85mm
%\epsfbox{figure4.eps}
%\vspace{0.2cm}
%\caption{Stationary solution for $\phi =20$, $L=80$ and $v_s=0.1$. $J$
%evolves towards $J\approx 6.2$ greater than $J_c=5.77$. The maximum
%electric field, $E_+ \approx 26.75$, is reached at $R_b \approx 35.12$.
%Field values at the boundaries are $E(r_c)=1.25$ and $E(r_a)=0.14$.}
%\label{fi20}
%\end{figure}
%\end{center}
%

We shall now describe the more interesting self-oscillations in Regime III,
starting with a bias interval of time-periodic oscillations. Stationary
solutions and their stability properties will be described elsewhere.

\subsection{Time-periodic oscillations for $0.35<\phi< 0.5$ and $v_s=0$}

\begin{center}
\begin{figure}[ht]
\epsfxsize=80mm
\epsfbox{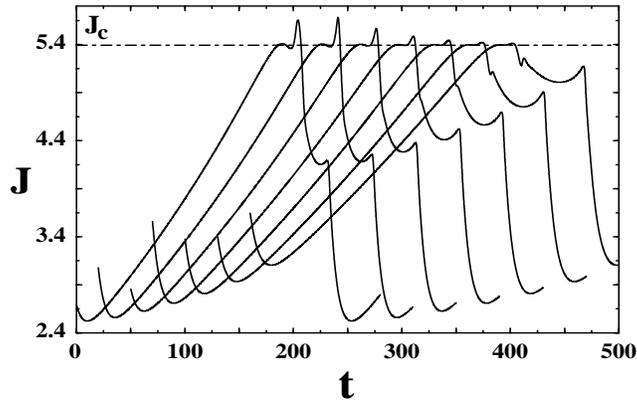}
\vspace{0.7cm}
\caption{Current vs time during one period of the self-oscillations for
$\phi\in (0.35,0.5)$. Values of $\phi$ are: 0.36, 0.38, 0.4, 0.42, 0.44,
0.46 and 0.48, depicted from left to right, and the critical current at
which a new wave is nucleated is $J_c=5.39$.}
\label{valleys}
\end{figure}
\end{center}

Fig.\ \ref{valleys} shows one period of $J(t)$ for different bias values in
this interval. The electric field profile $E(r,t)$ consists of a single
triangular pulse traveling towards the anode when $J(t)$ is increasing.
When $J(t)$ decreases, one triangular pulse disappears and a new one appears
at the cathode. The gradual increase of the current when there is only one
pulse in the sample lasts longer than the drop to low current values.
Notice that the current drop lasts the same for all bias values, while the
stage of current growth increases with $\phi$.

%
%\vspace{-0.5cm}
%\begin{center}
%\begin{figure}[ht]
%\epsfxsize=80mm
%\epsfbox{figure5.eps}
%\vspace{0.7cm}
%\caption{Current vs time during one period of the self-oscillations for
%$\phi\in (0.35,0.5)$. Values of $\phi$ are: 0.36, 0.38, 0.4, 0.42, 0.44,
%0.46 and 0.48, depicted from left to right, and the critical current at
%which a new wave is nucleated is $J_c=5.39$.}
%\label{valleys}
%\end{figure}
%\end{center}
%

This situation is remarkably different from that for the one-dimensional
geometry corresponding to
parallel planar contacts: the current signal is flat when there is only a
single pulse far from the contacts and the stages of current increase and
drop are very short \cite{hig92}. Other noticeable features in Fig.\
\ref{valleys} are: (i) a new wave is nucleated as $J$ surpasses a critical
value $J_c$ (bias independent); (ii) the current overshoot above $J_c$
decreases as $\phi$ increases and (iii) there is a second local maximum of
the current and the width of the region between the two local maxima
increases with $\phi$. Let us explain in more detail the field profiles
corresponding to these stages of the self-oscillation.

\begin{figure}[hb]
\begin{center}
\epsfxsize=80mm
\epsfbox{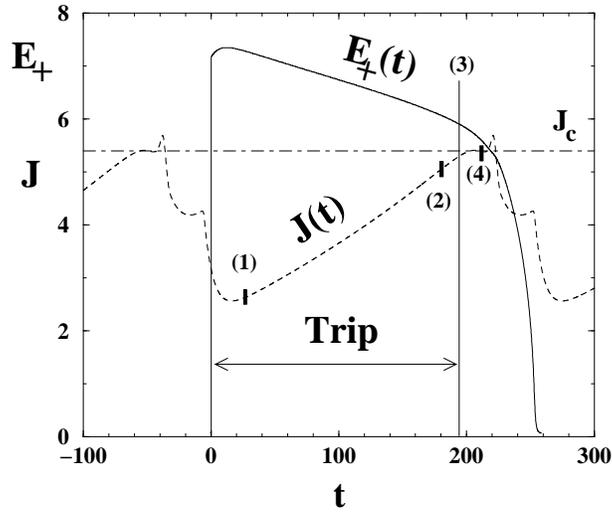}
\vspace{0.3cm}
\caption{Evolution of $J(t)$ and $E_+(t)$ during one oscillation period
for $\phi =0.38$ and $v_s=0$, during the time interval $0< t < \Delta
t\approx 200$, where there is a single pulse in the sample. Marked times
are: (1) 30, (2) 180, (3) 195, (4) 215.}
\label{Jtrip}
\end{center}
\end{figure}

\begin{center}
\begin{figure}[hb]
\epsfxsize=80mm
\epsfbox{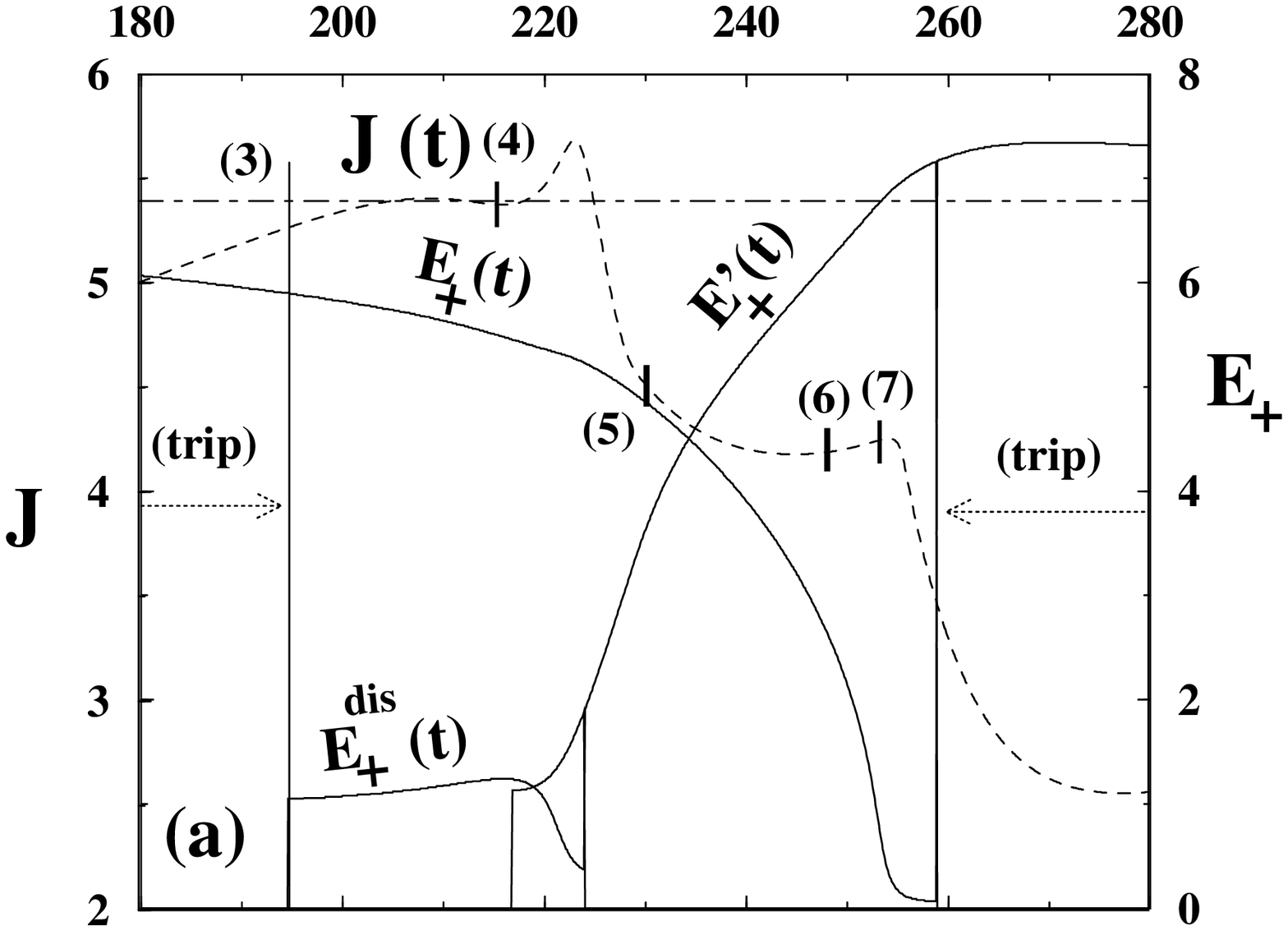}
\epsfxsize=80mm
\vspace{-.4cm}
\mbox{\hspace{1.5mm} \epsfbox{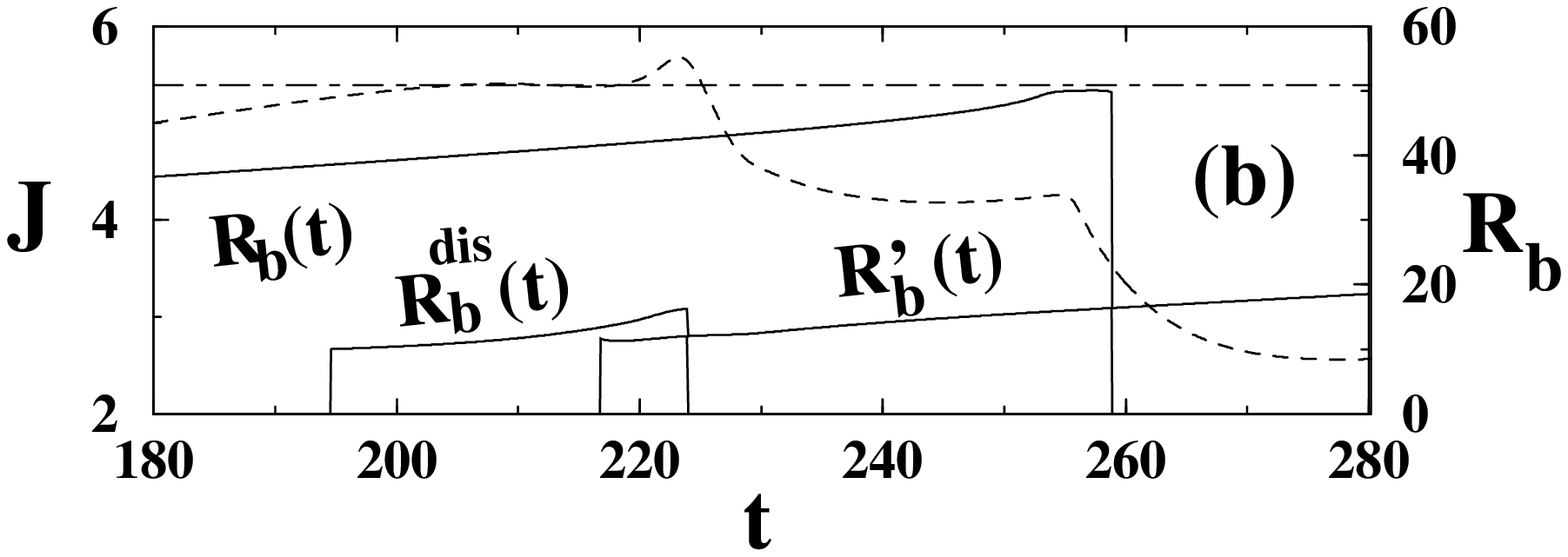}}
\vspace{0.7cm}
\caption{Evolution of $J(t)$ and $E_+(t)$ during one oscillation period
for $\phi =0.38$ and $v_s=0$. (a) Stage in which there are multiple pulses.
(b) $R_b(t)$ during the multi-pulse stage. $R_b^{dis}(t)$ and $R'_b(t)$ 
are the locations of the maximum fields of the transient field
disturbance and of the new pulse, respectively. Marked times are: (3)
195, (4) 215, (5) 230, (6) 248, and (7) 253.}
\label{Jrelief}
\end{figure}
\end{center}

\begin{center}
\begin{figure}[hb]
\epsfxsize=80mm
\epsfbox{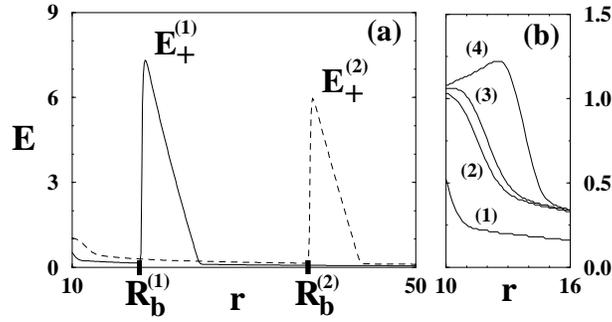}
\vspace{0.3cm}
\caption{(a) Electric field profile during the the single-pulse stage at
times (1) and (2) of Fig.\ \ref{Jtrip}. (b) Details of the field profile
near the cathode: the slope at $r_c$ increases with $J$; $dE(r_c)/dr=0$ for
$J=J_c \approx 5.24$ at time (3). }
\label{trip2}
\end{figure}
\end{center}

\begin{figure}[hb]
\begin{center}
\epsfxsize=80mm
\epsfbox{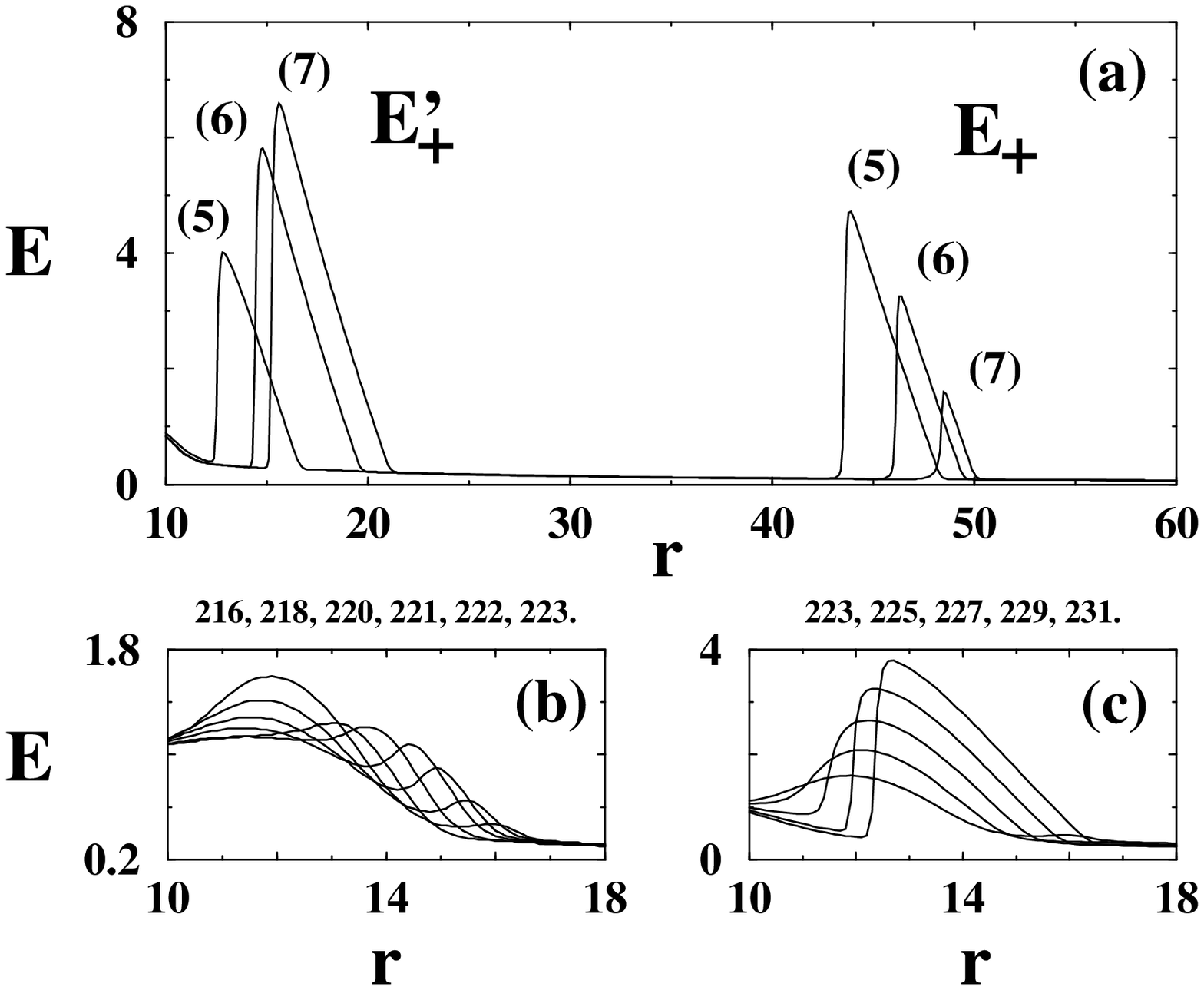}
\vspace{0.3cm}
\caption{(a) Electric field profiles during the multi-pulse stage at
times marked as (5), (6) and (7) in Fig.\ \ref{Jtrip} for $\phi =0.38$ and
$v_s=0$. (b) Details of the unsuccessful attempt at shedding one pulse
from the cathode for $J<J_c$.
(c) Successful nucleation of the new pulse after time (4).}
\label{relief}
\end{center}
\end{figure}

%\vspace{-.5cm}
%\begin{figure}[hb]
%\begin{center}
%\epsfxsize=80mm
%\epsfbox{figure6.eps}
%\vspace{0.3cm}
%\caption{Evolution of $J(t)$ and $E_+(t)$ during one oscillation period
%for $\phi =0.38$ and $v_s=0$, during the time interval $0< t < \Delta
%t\approx 200$, where there is a single pulse in the sample. Marked times
%are: (1) 30, (2) 180, (3) 195, (4) 215.}
%\label{Jtrip}
%\end{center}
%\end{figure}
%
Fig.\ \ref{Jtrip} and Fig.\ \ref{Jrelief} show details of the current
signal during one period of the self-oscillation for $\phi=0.38$ and
$v_s=0$. Also shown are the maxima of the electric field the different pulses
that appear in this time interval. From $t=0$ to $t=195$, there is a
single pulse moving towards $r_a$. This wave is roughly a straight
triangle of height and width $E_+(t)$, which is the maximum of the field
inside the wave at time $t$. The back of the pulse can be approximated by
a shock wave located at $R_b(t)$. As we describe below, self-oscillations
in the bias interval considered here involves creation of new pulses and
of transient field disturbances at the cathode. In Fig.\ \ref{Jrelief},
$R'_b(t)$ and $R^{dis}_b(t)$ are the locations of the maximum field of a
new pulse and of a transient pulse-like field disturbance (both shed at
the cathode), respectively. Fig.\ \ref{trip2}(a) shows the electric field
profile at the times marked (1) and (2) in Fig.\ \ref{Jtrip}. Fig.\
\ref{relief}(a) depicts the electric field near the cathode. A new pulse
is shed from the cathode when $J$ surpasses the critical value $J_c$ at
time (4) in Fig.\ \ref{Jtrip}. Between times (3) and (4) (with $J<J_c$),
a field disturbance is shed from the cathode and it shrinks rapidly as it
advances. Eventually as the first pulse disappears (slightly after the
time corresponding to the second local maximum of the current), the new
pulse shed at time (4) becomes the only one in the sample. A new
oscillation period starts then. Fig.\ \ref{relief} shows details of the
field profiles when there is more than one pulse in the sample.

%\vspace{-.4cm}
%\begin{center}
%\begin{figure}[hb]
%\epsfxsize=80mm
%\epsfbox{figure7a.eps}
%\epsfxsize=80mm
%\vspace{-.4cm}
%\mbox{\hspace{1.5mm} \epsfbox{figure7b.eps}}
%\vspace{0.7cm}
%\caption{Evolution of $J(t)$ and $E_+(t)$ during one oscillation period
%for $\phi =0.38$ and $v_s=0$. (a) Stage in which there are multiple pulses.
%(b) $R_b(t)$ during the multi-pulse stage. Marked times are: (3) 195, (4)
%215, (5) 230, (6) 248, and (7) 253.}
%\label{Jrelief}
%\end{figure}
%\end{center}
%

%
%\vspace{-2.4cm}
%\begin{center}
%\begin{figure}[hb]
%\epsfxsize=80mm
%\epsfbox{figure8.eps}
%\vspace{0.3cm}
%\caption{(a) Electric field profile during the the single-pulse stage at
%times (1) and (2) of Fig.\ \ref{Jtrip}. (b) Details of the field profile
%near the cathode: the slope at $r_c$ increases with $J$; $dE(r_c)/dr=0$ for
%$J=J_c \approx 5.24$ at time (3). }
%\label{trip2}
%\end{figure}
%\end{center}
%

%
%\begin{figure}[hb]
%\begin{center}
%\epsfxsize=80mm
%\epsfbox{figure9.eps}
%\vspace{0.3cm}
%\caption{(a) Electric field profiles during the multi-pulse stage at
%times marked as (5), (6) and (7) in Fig.\ \ref{Jtrip}.
%(b) Details of the unsuccessful attempt at shedding one pulse from the
%cathode for $J<J_c$.
%(c) Successful nucleation of the new pulse after time (4).}
%\label{relief}
%\end{center}
%\end{figure}

\subsection{Aperiodic current self-oscillations}
\begin{center}
\begin{figure}[ht]
\epsfxsize=80mm
\epsfbox{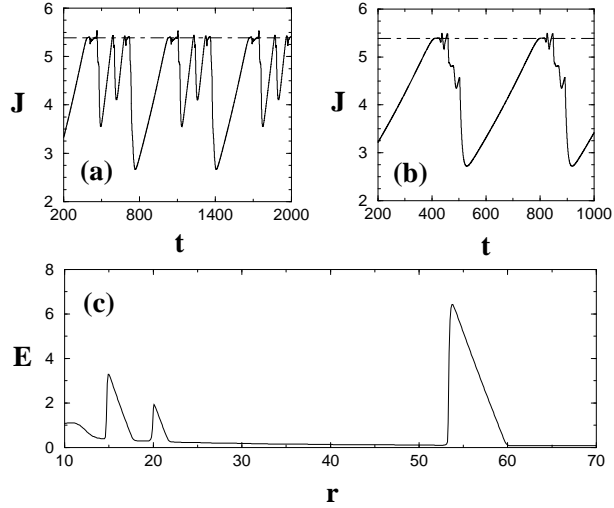}
\vspace{0.3cm}
\caption{(a) Complex current signal $J(t)$ for $\phi=0.5$ and a time
interval $\Delta t \approx 650$. (b) Same for $\phi=0.52$ and $\Delta t
\approx 400$. (c) Three-pulse electric field profile for $\phi=0.52$ and
$\Delta t\approx 400$. In all cases, $v_s=0$. In (a) and (b), the
horizontal line marks the critical current $J_c$. }
\label{multidipole}
\end{figure}
\end{center}

For biases $\phi\geq 0.5$, periodic self-oscillations of the current
alternate with voltage ranges of aperiodic oscillations. Fig.\
\ref{multidipole} shows that the corresponding current signals may be
rather complex, with several maxima and current overshoots above $J_c$.
%
%\vspace{-0.3cm}
%\begin{center}
%\begin{figure}[ht]
%\epsfxsize=80mm
%\epsfbox{figure10.eps}
%\vspace{0.3cm}
%\caption{(a) Complex current signal $J(t)$ for $\phi=0.5$ and a time
%interval $\Delta t \approx 650$. (b) Same for $\phi=0.52$ and $\Delta t
%\approx 400$. (c) Three-pulse electric field profile. In all cases,
%$v_s=0$. }
%\label{multidipole}
%\end{figure}
%\end{center}
%\vspace{-0.75cm}

\begin{center}
\begin{figure}[ht]
\epsfxsize=80mm
\epsfbox{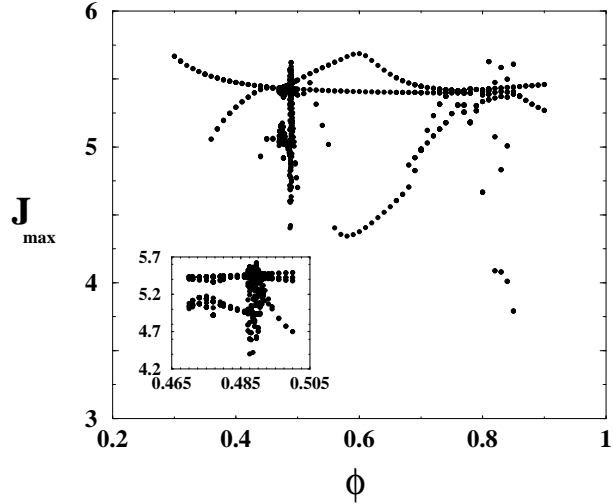}
\vspace{0.3cm}
\caption{Poincar\'e diagram depicting current maxima vs bias illustrating
loss of periodic oscillations in narrow bias intervals. Parameters are
$r_c=10$, $r_a=50$ and $v_s=0$. }
\label{poincare}
\end{figure}
\end{center}

Typically several pulses are present during different time intervals of
the self-oscillation. The leftmost pulse may shrink as the second pulse
increases, or it may reach it and coalesce with that pulse. Meanwhile the
rightmost pulse may shrink or reach the anode. Pulses may be shed from the
cathode or nucleate inside the sample. Resulting current signals may even
be apparently chaotic. A detailed study of all cases that are possible
depending on the bias will not be attempted here. Fig.\ \ref{poincare}
depicts the local maxima of the current as a function of bias. Loss of
periodicity at narrow bias intervals is apparent.
%
%\vspace{-0.3cm}
%\begin{center}
%\begin{figure}[ht]
%\epsfxsize=80mm
%\epsfbox{figure11.eps}
%\vspace{0.3cm}
%\caption{Poincar\'e diagram depicting current maxima vs bias illustrating
%loss of periodic oscillations in narrow bias intervals. Parameters are
%$r_c=10$, $r_a=50$ and $v_s=0$. }
%\label{poincare}
%\end{figure}
%\end{center}

\begin{center}
\begin{figure}[ht]
\epsfxsize=85mm
\epsfbox{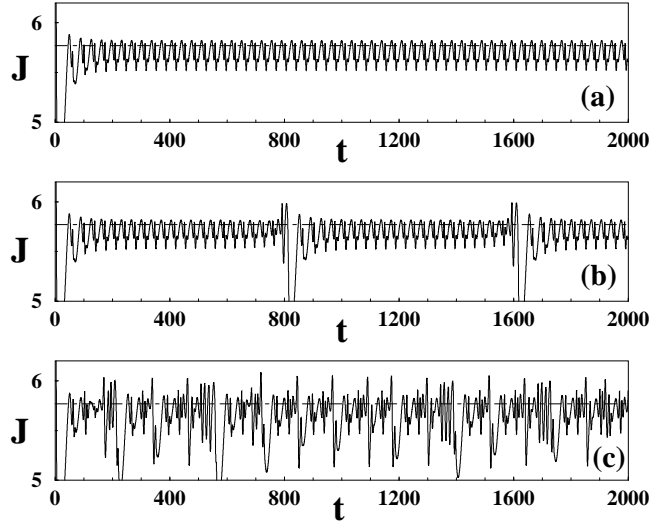}
\vspace{0.3cm}
\caption{Complex current signals for $v_s= 0.1$, $r_c=10$, $r_a=50$ and
several bias values: (a) $\phi = 0.41$ (periodic signal), (b) $\phi = 0.411$
(periodic signal with a longer period due to current bursts), and (c)
$\phi = 0.42$ (aperiodic signal). In all cases, $J(t)$ oscillates about $J_c=
5.77$ with an approximate amplitude of 0.5.  }
\label{0.1}
\end{figure}
\end{center}

Comparing the case $v_s>0$ to that with $v_s=0$, we observe that the
pulses move faster, the oscillations have smaller amplitudes and it is
easier for several pulses to coexist when $v_s>0$. This may result in more
complex shapes of the current signal, as shown in Fig.\ \ref{0.1}. The current
signal is periodic in Figures \ref{0.1}(a) and \ref{0.1}(b) although the
period is longer in the latter due to separated current bursts. The current
signal in Fig.~\ref{0.1}(c) is aperiodic. 
%\vspace{0.5cm}
%\begin{center}
%\begin{figure}[ht]
%\epsfxsize=85mm
%\epsfbox{figure12.eps}
%\vspace{0.3cm}
%\caption{Complex current signals for $v_s= 0.1$, $r_c=10$, $r_a=90$ and
%several bias values. $J(t)$ oscillates about $J_c= 5.77$ with an
%approximate amplitude of 0.5.}
%\label{0.1}
%\end{figure}
%\end{center}
%\vspace{-0.75cm}

\section{Interpretation of the numerical results}
\label{sec:4}
In order to understand the shape of $J(t)$ in the time periodic regime and the
fact that the pulses may vanish before reaching the anode, we present in
this section a qualitative description of the asymptotic oscillatory solution
for $r_a \gg r_c \gg 1$. In this case there is a wide range of voltages for
which the pulses are detached from the contacts during most of their evolution
time. A full analysis of the pulse dynamics should include descriptions of the
evolution of a pulse far from the contacts and of the generation of new pulses
at the injecting contact. The latter process is essentially as for the planar
case \cite{hig92} when $r_c \gg 1$, because the effect of the geometrical
divergence is then negligible around the cathode. Analysis shows that, roughly,
a pulse is shed when the current increases to $J=J_c \approx j_c r_c$, and then
the current decreases while the new pulse grows and separates from the cathode.
These results are in line with the numerical simulations of the previous
section.

In the remainder of this section we focus on the evolution of a pulse detached
from the cathode. As in the planar case, such a pulse is a straight triangle
made of a trailing edge which is a shock and a leading ramp which is a region
depleted of electrons \cite{hig92}.

Consider first the region outside of the pulse. Time and space derivatives can
be neglected in (\ref{7}) for this region, which covers most of the sample,
leading to the appoximate solution $v(E) = J/r$, which implies
$E(r,t) = E_1(J/r)$,
where $E_1(j)<E_2(j)$ are the two solutions of $v(E)=j$ for $v_s< j < v_M$.
If $J\ll r$, the first branch of $v(E)$ is linear, and we have
$E(r,t)\approx J/r$. The area under this stationary field profile is
$\Phi_{out} = \int_{r_{c}}^{r_{a}} E_1(J/r)\, dr\approx J \ln(r_a/r_c)$.

The speed of trailing edge of the pulse, at $r=R_b(t)$, is given by the equal
area rule for a shock rising the field from $E_-$ to $E_+$; i.e.
$V(E_+,E_-)\equiv \int_{E_-}^{E_+} v(E)\, dE/(E_+ - E_-)$. For
large pulses, $E_+\gg 1$ and the field immediately at the left of the shock is
$E_- < E_M = O(1)$, so that we may approximate
\begin{eqnarray}
V(E_+,E_-)= v_s + {\int_{E_-}^{E_+} [v(E)-v_s]\, dE \over E_+ - E_-}
% \nonumber\\
\sim v_s + {C \over E_+}
\label{Eq2}
\end{eqnarray}
with
\begin{eqnarray}
C= \int_{E_1(v_s)}^{\infty} [v(E)-v_s]\, dE. \label{Eq3}
\nonumber
\end{eqnarray}
Here we have used that $V\sim v_s$ and $E_-\sim E_1(v_s)$ as $E_+\gg 1$.
If $v_s=0$, then $C= \pi/4$ and $V(E_+,E_-)\sim \pi/(4E_+)$. In this case, the
trailing front velocity is small and small waves move faster than large
ones. If $v_s>0$, then the waves move at a speed close to the saturation
speed $v_s$.

The electron density at the leading ramp of the pulse
is almost zero, so that the field obeys the Poisson equation
$1+r^{-1}\, \partial (rE)/\partial r = n \approx 0$, whose solution is
\begin{eqnarray}
E(r,t) = {r^{2}_w(t) - r^{2} \over 2r} \approx r_w - r
\label{Leadingfront}
\end{eqnarray}
in the pulse. Here the constant of integration $r_w(t)$ is the
intersection between the prolongation of the ramp and the $r$ axis, and use
has been made of the condition that the width of the pulse
$W = r_w - R_b \approx E_+$ satisfies $r_w \gg W \gg 1$ to simplify the
result. The area of the triangular pulse is
$\Phi_{in} \approx E_+^2/2$, and therefore the bias condition (\ref{8}) becomes
\begin{eqnarray}
\Phi= \Phi_{in}+\Phi_{out} \sim {E_+^{2}\over 2} + J\, \ln\left( {r_a\over
r_c}\right)\,. \label{Sys1}
\end{eqnarray}

The speed of the ramp, $dr_w/dt$, coincides with the speed of the
electrons immediately ahead of the pulse
\begin{eqnarray}
{dr_w\over dt} = {J\over r_w} \,.
\label{Eq5}
\end{eqnarray}
This result can also be obtained by inserting (\ref{Leadingfront}) and $n=0$
into (\ref{7}). Equations (\ref{Sys1}) and (\ref{Eq5}), along with
\begin{eqnarray}
{d(r_w - E_+) \over dt} = v_s + {C \over E_+}
\label{TS}
\end{eqnarray}
for the speed of the trailing shock (at $R_b \approx r_w-E_+$), suffice to
determine the time evolution of $J$, $r_w$ and $E_+$ from a given initial
state. Equation~(\ref{TS}) can be rewritten as
\begin{eqnarray}
\frac{dJ}{dt} = {1 \over \ln(r_a/r_c)} \left\{C - \left({J\over r_{w}} -
v_s\right) \times\right. \nonumber \hspace{1cm} \\
\left. \sqrt{2\left[\Phi - J\ln\left({r_a\over r_c}\right)\right]}
\right\}\,,\label{Sys3}
\end{eqnarray}
where use has been made of (\ref{Sys1}) and (\ref{Eq5}).

\begin{center}
\begin{figure}[ht]
\epsfxsize=86mm
\epsfbox{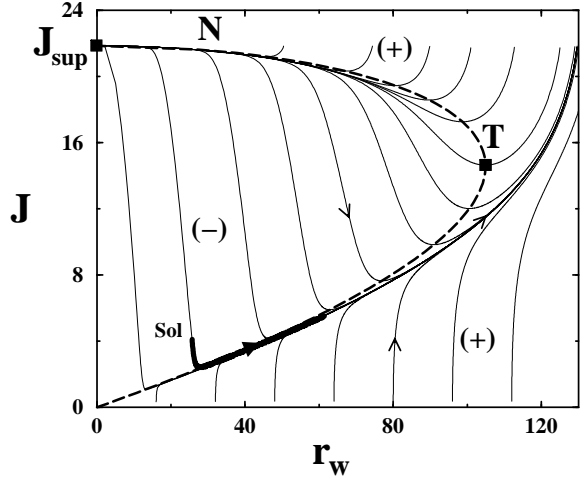}
\caption{Phase plane $(r_w,J)$ showing the nullcline $d J/d r_w=0$ (dash line)
and the turning point for $v_s=0$ and a bias $\phi=0.6$. The thick line
represents the trajectory of the solution for initial data $J(0)=4.1$ and
$r_w = 26$, until $J_c\approx 5.4$ is reached.  }
\label{phase}
\end{figure}
\end{center}

The problem can be further simplified noticing that the speeds of the shock and
of the ramp are nearly equal to each other during most of the pulse lifetime,
\begin{eqnarray}
v_s +{C \over E_+} = {J\over r_w}.
\label{Eq55}
\end{eqnarray}
This occurs because the width of the pulse is small compared with the total
distance it travels in the sample. Then a mismatch of the two speeds would
lead either to the disappearance of the pulse or to its growth above the
maximum size allowed by the bias in a time of order $E_+/\max (V, J/r_w)$.
This time is short compared with the pulse lifetime. An algebraic relation
between $J$ and $r_w$ can be obtained eliminating $E_+$ between (\ref{Sys1})
and (\ref{Eq55}). This is plotted as a dashed curve in Fig.\ \ref{phase},
which is the phase plane of (\ref{Eq5}) and (\ref{Sys3}) for a case with
$v_s=0$. As can be seen, the trajectories tend rapidly to the lower branch of
the dashed curve and then rise along this branch obeying (\ref{Eq5}) until
either: (i) $r_w$ reaches its maximum possible value, corresponding to the
turning point $T$, or (ii) $J$ reaches the critical value $J_c$ for nucleation
of a new pulse, whichever happens first.
%
%\vspace{-0.5cm}
%\begin{center}
%\begin{figure}[ht]
%\epsfxsize=86mm
%\epsfbox{figure13.eps}
%\caption{Phase plane $(r_w,J)$ showing the nullcline $d J/d r_w=0$ (dash line)
%and the turning point for $v_s=0$ and a bias $\phi=0.6$. The thick line
%represents the trajectory of the solution for initial data $J(0)=4.1$ and
%$r_w = 26$, until $J_c\approx 5.4$ is reached.  }
%\label{phase}
%\end{figure}
%\end{center}
%\vspace{-0.75cm}

\begin{figure}[ht]
\begin{center}
\epsfxsize=86mm
\epsfbox{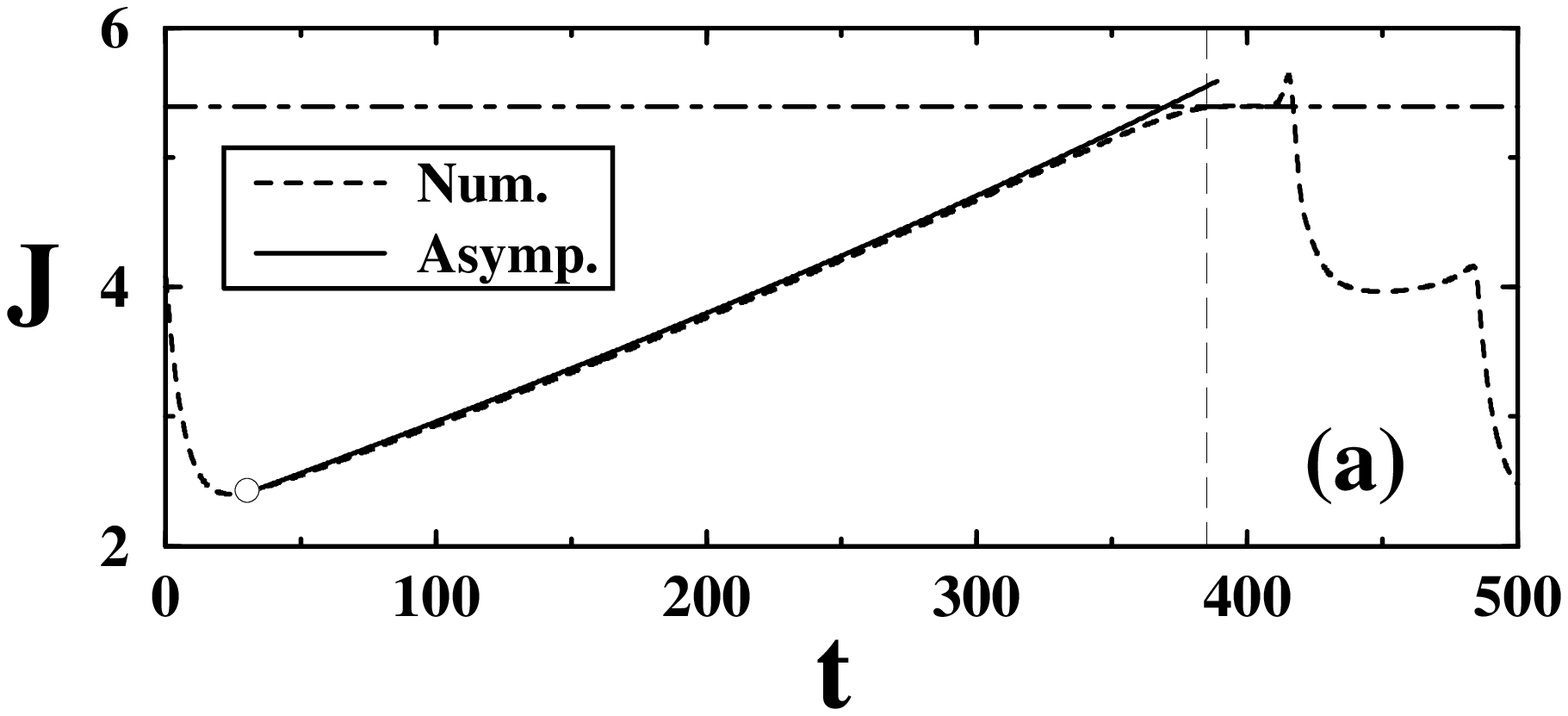}\\
\vspace{-1.9cm}
\epsfxsize=86mm
\epsfbox{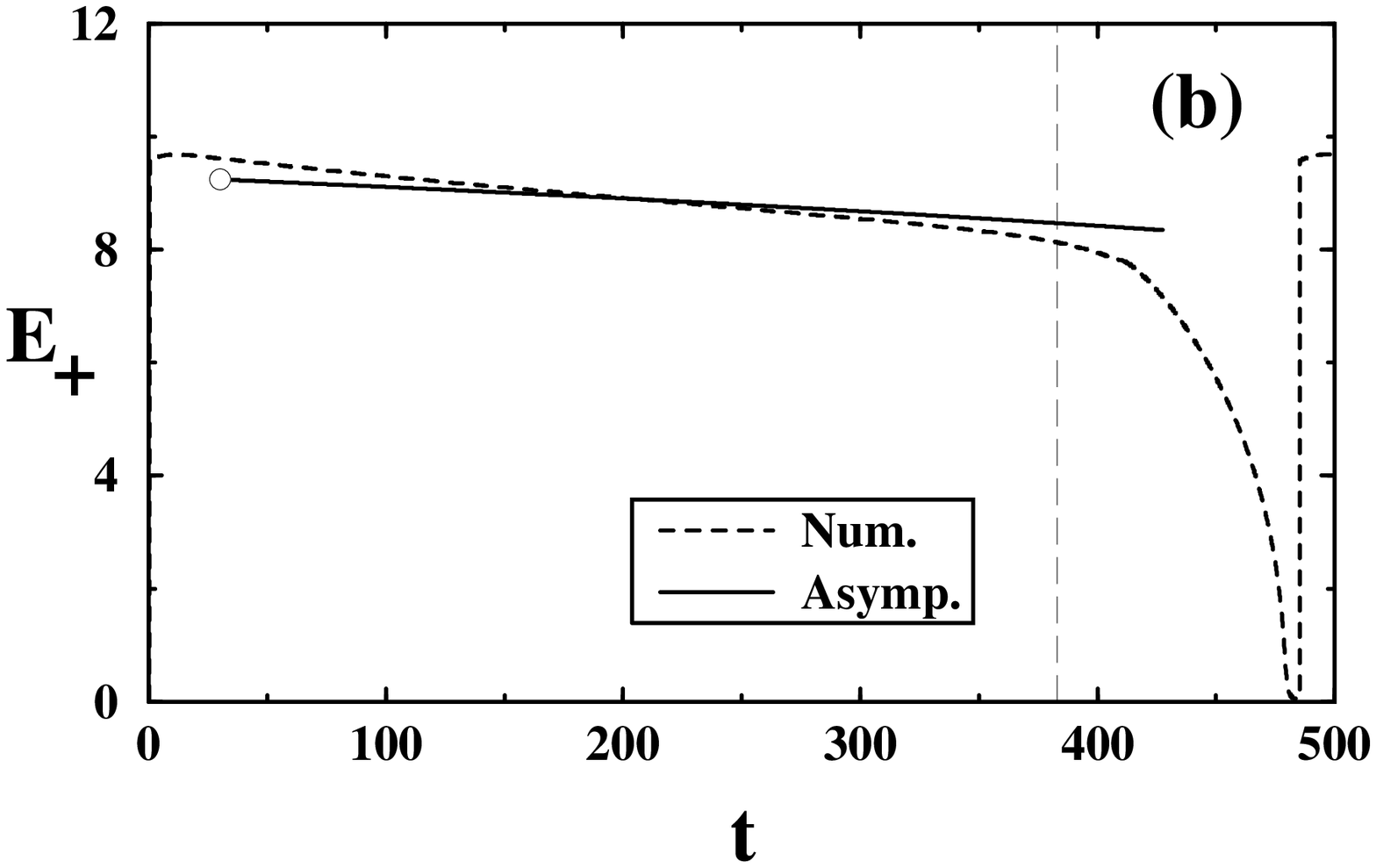}\\
\vspace{-3.6cm}
\epsfxsize=86mm
\epsfbox{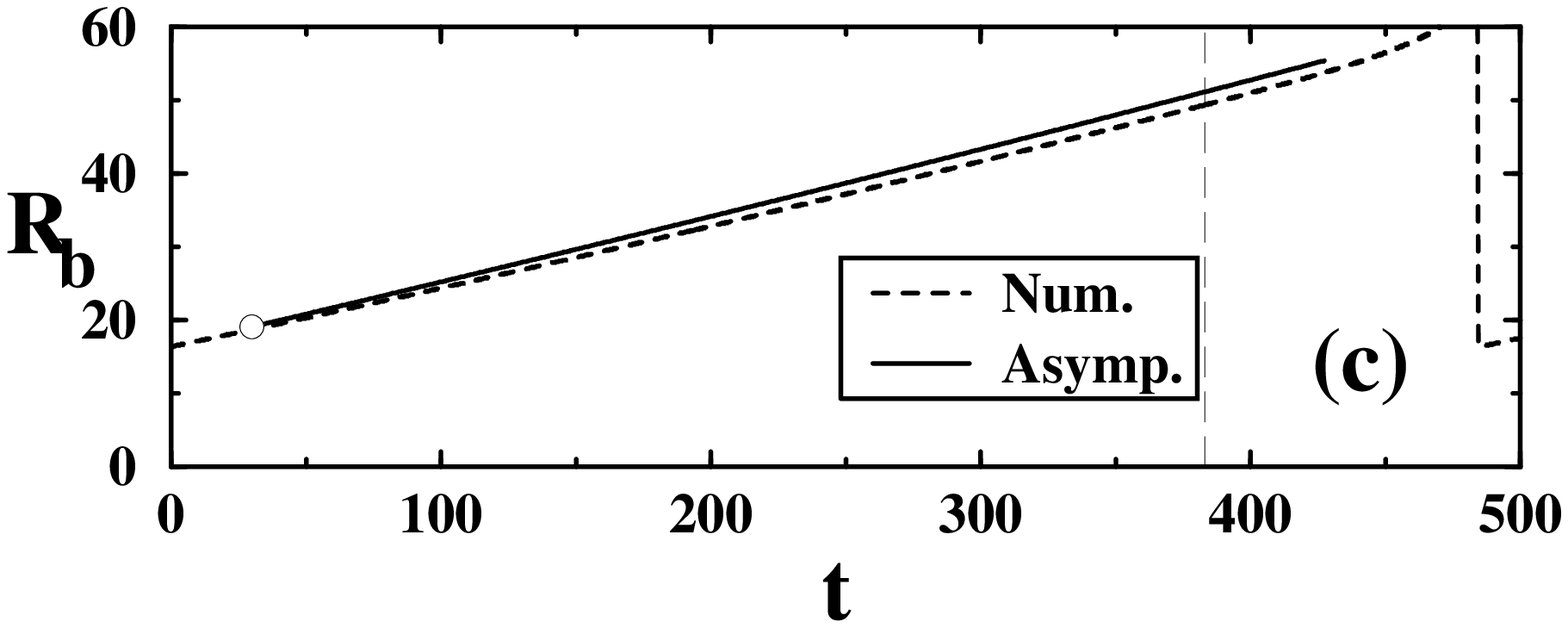}\\
\vspace{-3.7cm}\hspace{-.5cm}
\epsfxsize=86mm
\epsfbox{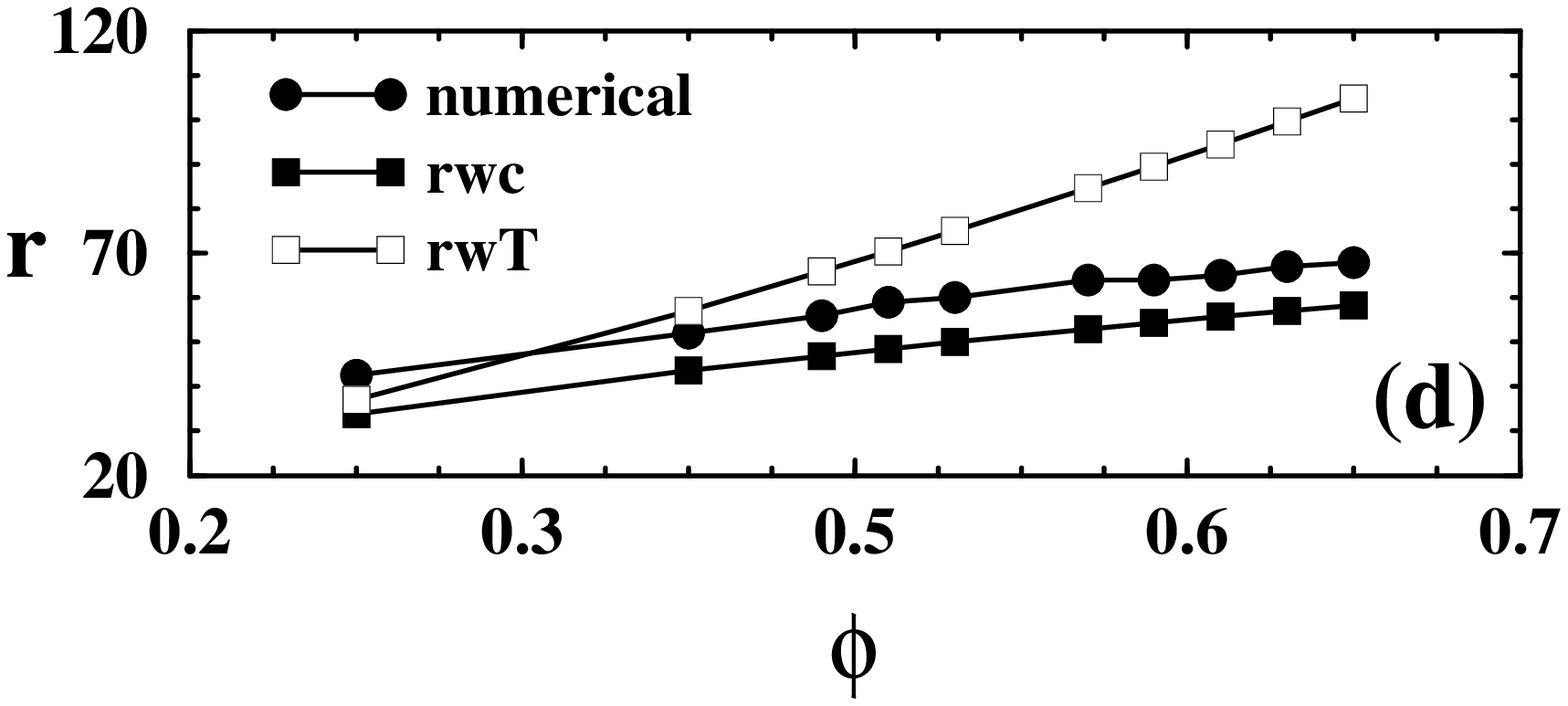}
\vspace{.3cm}
\caption{Comparison of the asymptotic expressions for $J$, $E_+$ and $R_b$
with the results of direct numerical solution starting with $J(t_0=30)=
J_{min} \approx 2.4$. Parameter values are $r_c=10$, $r_a=90$, $\phi= 0.6$
and $v_s=0$. (a) $J(t)$ (the horizontal line marks the critical current
$J_c$). (b) $E_+(t)$. (c) $R_b(t)$. (d) Maximum radius as
a function of bias. The numerically calculated maximum radius for the back
of the pulse is compared to $r_{wT}$ and $r_{wc}$. The difference of the
latter with the numerically calculated $R_b$ is the pulse width.}
\label{compare}
\end{center}
\end{figure}

When $v_s=0$ the turning point is
\begin{eqnarray}
J_T= {2\Phi\over 3\ln\left({r_a\over r_c}\right)}\,,\quad
r_{wT} = {4\over \pi\ln\left({r_a\over r_c}\right)}\, \left({2\Phi\over 3}
\right)^{{3\over 2}}\,. \label{Eq6}
\end{eqnarray}
Then maximum radius of the pulse is either $r_{wT}$ or
\begin{eqnarray}
r_{wc} = {4J_{c}\over \pi}\,\sqrt{2\left[\Phi-\ln\left({r_a\over
r_c}\right)\right]}\,, \label{Eq7}
\end{eqnarray}
if $J_{c}< J_T$. The time dependence of $J$ from its minimum value
during an oscillation period can be obtained by integrating (\ref{Eq5}) along
the dashed line (nullcline) of Fig.\ \ref{phase}. In the present case of
$v_s=0$, Eq.~(\ref{Eq5}) can be written as
$[\Phi - (3J/2)\ln(r_a/r_c)]\, dJ/dt = \pi^2/32$, with the help of
(\ref{Sys1}) and (\ref{Eq55}).
Upon integrating this equation,
\begin{eqnarray}
J(t) = J_T\, \left(1-\sqrt{\left(1-{J_{min}\over J_{T}}\right)^2 - {t\over
t_T}}\right)\,, \label{Eq8}\\
t_{T} = {32\Phi^2\over 3\pi^2\ln\left({r_a\over r_c}\right)}\,. \label{Eq9}
\end{eqnarray}
We have used as initial condition the minimum value of the current during
one oscillation period, $J(0)=J_{min}$. A comparison of this approximation
to a direct numerical solution of the whole problem is shown in Fig.\
\ref{compare}.

%$\,$
%\vspace{-3.43cm}
%\begin{figure}[ht]
%\begin{center}
%\epsfxsize=86mm
%\epsfbox{figure14a.eps}\\
%\vspace{-1.9cm}
%\epsfxsize=86mm
%\epsfbox{figure14b.eps}\\
%\vspace{-3.6cm}
%\epsfxsize=86mm
%\epsfbox{figure14c.eps}\\
%\vspace{-3.7cm}\hspace{-.5cm}
%\epsfxsize=86mm
%\epsfbox{figure14d.eps}
%\vspace{.3cm}
%\caption{Comparison of the asymptotic expressions for $J$, $E_+$ and $R_b$
%with the results of direct numerical solution starting with $J(t_0=30)=
%J_{min} \approx 2.4$. Parameter values are $r_c=10$, $r_a=90$, $\phi= 0.6$
%and $v_s=0$. (a) $J(t)$. (b) $E_+(t)$. (c) $R_b(t)$. (d) Maximum radius as
%a function of bias. The numerically calculated maximum radius for the back
%of the pulse is compared to $r_{wT}$ and $r_{wc}$. The difference of the
%latter with the numerically calculated $R_b$ is the pulse width.}
%\label{compare}
%\end{center}
%\end{figure}

For shorter samples such that $r_a <r_{wc}$ and large bias, the pulse
reaches the anode before its maximal radius. Then the situation resembles
the one-dimensional Gunn effect with large voltage: a pulse reaches
the anode before a new pulse is nucleated at the cathode \cite{hig92}. If
we relax the assumption $v_s=0$, we find that the pulse speed is faster
according to (\ref{Eq2}). As in the case of $v_s=0$, there is also a maximum
radius of the pulse (for large enough samples). In fact, the nullcline
$d J/d r_w=0$ has a turning point which can be found by solving
\begin{eqnarray}
{v_s\over C}\, z^3 + {3\over 2}\, z^2 - \Phi =0,\label{Eq10}
\end{eqnarray}
for $z>0$ as a function of $\Phi$. Then $J_T$ and $r_{wT}$ are given by
\begin{eqnarray}
z= E_+ = \sqrt{2\, \left[\Phi - J_T\, \ln\left({r_a\over r_c}\right)
\right]},\label{Eq11}\\
r_{wT} = {J_T\, z\over C}\,. \label{Eq12}
\end{eqnarray}
An approximate solution of the system (\ref{Eq5}) and (\ref{Sys3}) can be
found as before.
The maximum radius can be found by solving the simplified problem
with initial condition $J_{min}$ and calculating the time that
$J$ needs to reach $J_c$. In general, the pulse reaches either its
maximum radius or the anode (depending on $L$ and $\phi$) much earlier
than in the case of $v_s=0$. This leads to small-amplitude current
oscillations which may be rather irregular because $J$ may be above
$J_c$ quite often, thereby producing new pulses near the cathode. See
Fig.\ \ref{0.1}.

\section{Conclusions}
\label{sec:5}
We have studied numerically repeated generation and motion of axisymmetric
pulses in a two-dimensional n-GaAs sample with a Corbino geometry. The
field inside these pulses decreases as they advance and expand, so as to
compensate their larger extension. Simultaneously, the current increases
until a critical value is reached and a new pulse is triggered at the
central point contact. The current signal presents different patterns
depending on the applied dc voltage bias. Just above the onset for
self-oscillations, their amplitude is small and the pulse dies off shortly
after it is generated. For larger voltages, self-oscillations have larger
amplitude and pulses may or may not reach the outer sample boundary
depending on the size thereof and bias. For sufficiently large samples, the
pulse radius cannot surpass a maximum value given by an approximate
analytical formula. Regions of aperiodic oscillations due to multi-pulse
dynamics are interspersed with more regular periodic oscillations.

\acknowledgements
This work has been supported by the Spanish DGES through grant
PB98-0142-C04.

\appendix
\section{Outline of the numerical method}
\label{appendix}
\begin{center}
\begin{figure}[ht]
\epsfxsize=80mm
\epsfbox{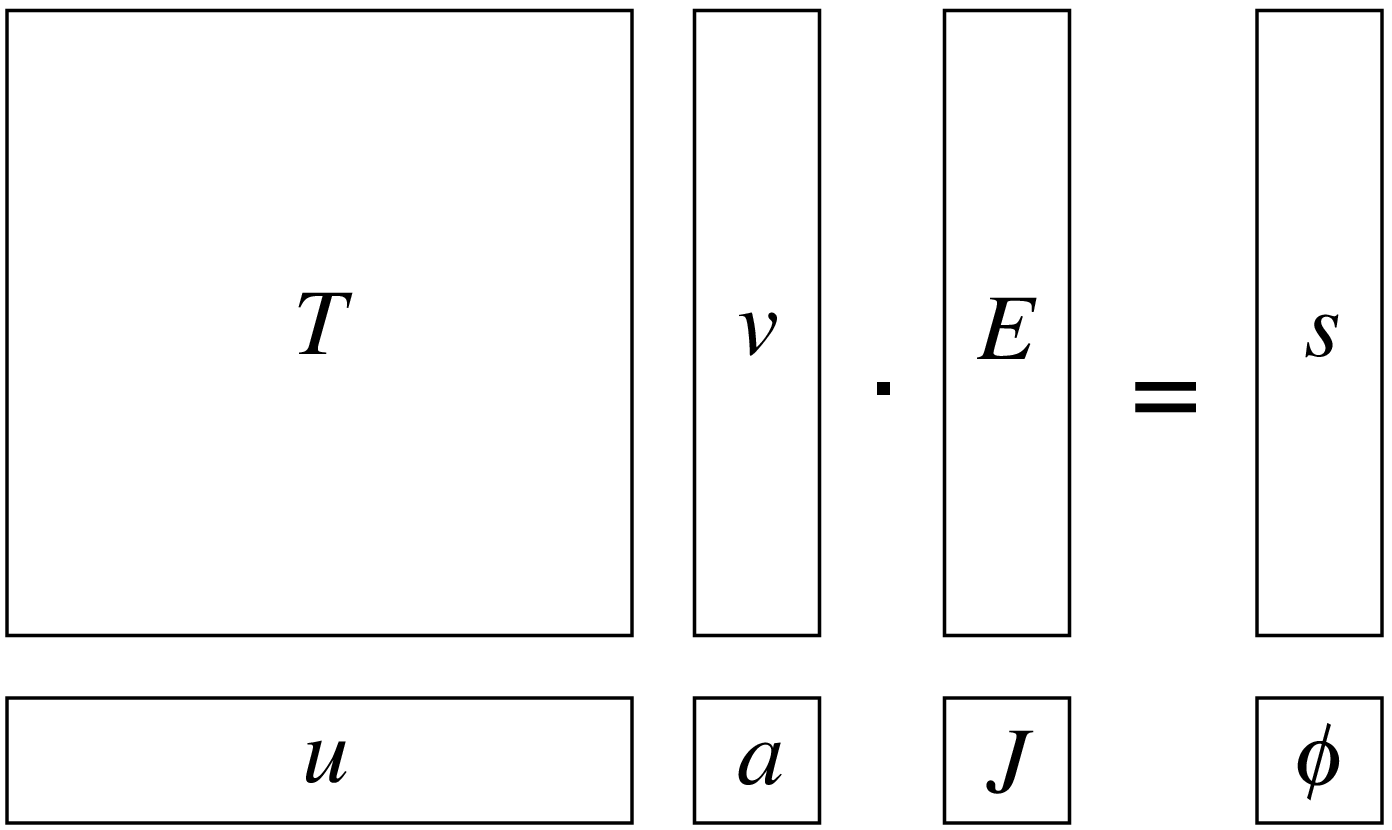}
\vspace{.3cm}
\caption{Block matrix formulation of our
numerical scheme to solve the equations for $E$
and $J$.}
\label{matrix}
\end{figure}
\end{center}

We have used an efficient numerical scheme for partial differential
equations with an integral constraint described and proved to converge in
Ref.\ \onlinecite{car01}. Radial derivatives are approximated by central
differences and a first-order implicit Euler method is used to integrate
the resulting differential equations in time. This procedure results in
having to solve a system of $(N+1)$ linear equations for the values of the
electric field and $J$ at time $t_{n+1}$ in terms of their previous values.
The block matrix formulation of this system is
shown in Fig.\ \ref{matrix}. There ${\bf T}$,
${\bf u}$ and
${\bf v}$ are a $N\times N$ tridiagonal matrix, a
$1\times N$ row vector and a $N\times 1$ column
vector, respectively. Our system is therefore
equivalent to
\begin{eqnarray}
{\bf T } \cdot {\bf E} \, + \, J {\bf v }  & = & {\bf s},
 \label{11} \\
{\bf u } \cdot {\bf E} \, + \, J a & = & \phi.
 \label{22}
\end{eqnarray}
This system can be efficiently solved by solving the following two systems
with the same tridiagonal matrix:
\begin{eqnarray}
{\bf{T}} \cdot {\bf{y}} = {\bf{s}},
\label{y}\\
{\bf{T}} \cdot {\bf{z}} = {\bf{v}}.
\label{z}
\end{eqnarray}
In terms of ${\bf y}$ and ${\bf z}$, we obtain
\begin{eqnarray}
{\bf E} = {\bf{y}} - J {\bf{z}}, \label{E=y-Jz}\\
J = \frac{{\bf u} \cdot {\bf y} - \phi}{{\bf u} \cdot  {\bf z} - a}.
\label{J=}
\end{eqnarray}
Thus we proceed by firstly obtaining the $LU$ factorization of ${\bf T}$ 
and secondly carrying out two back substitution processes to solve
(\ref{y}) and (\ref{z}). Then (\ref{J=}) and (\ref{E=y-Jz}) yield $J$ and
${\bf E}$, respectively.

%\newpage
\vspace{-.2cm}

\vspace{-0.4cm}

%\vspace{.2cm}

\vspace{-.5cm}

\vspace{-0.5cm}
%
%\vspace{-.5cm}
\vspace{-.4cm}
%
%\vspace{-2.4cm}

\vspace{-0.3cm}
%\vspace{-0.75cm}
%

\vspace{-0.3cm}

%\vspace{0.5cm}
%\vspace{-0.75cm}

%\vspace{-0.5cm}
%\vspace{-0.75cm}

$\,$
\vspace{-3.43cm}

%
%\listoffigures
%\end{multicols}
\end{document}